\documentstyle[12pt]{article}
\input epsf

\begin{document}
\rightline{EFI-94-21}
\rightline{hep-ph/9405404}

\begin{center}
\Large\bf \boldmath CKM Matrix Elements:  Magnitudes, Phases, Uncertainties
\footnote{Invited talk published in {\it Proceedings of the 2nd IFT Workshop
on Yukawa Couplings and the Origins of Mass}, Gainesville, FL, 11-13 February
1994, edited by P. Ramond (International Press, 1996), pp.~273--293.}
\unboldmath
\end{center}

\medskip
\centerline{Jonathan L. Rosner}
\centerline{\it Enrico Fermi Institute and Department of Physics}
\centerline{\it University of Chicago, Chicago, IL 60637}
\vskip 1cm

\centerline{\bf ABSTRACT}
\bigskip

\begin{quote}
Schemes for fermion masses should predict elements of the
Cabibbo-Kobayashi-Maskawa (CKM) matrix.  We review the freedom allowed by
present experiments and how the parameter space will shrink in the next few
years.  In addition to experiments which directly affect CKM parameters, we
discuss constraints arising from precise electroweak tests.
\end{quote}
\bigskip

\section{Introduction}

The present workshop is devoted to ways in which the bewildering pattern of
fermion masses and mixings might be understood.  The purpose of this talk is to
describe the allowed parameter space of mixings and how it may be expected to
shrink as a result of improved experiments and theory.  A more detailed account
may be found in Ref.~\cite{CKM}; we take the opportunity to update some of
the numbers presented there.  Some of the latest developments since this talk
was presented will be mentioned, but are not included in the fits to data. 

We set out the frameworks for the discussion in Section~\ref{sec:fra}. The
determination of CKM parameters is described in Section~\ref{sec:CKM}. A long
digression on the top quark is contained in Section~\ref{sec:top}. Electroweak
tests lead primarily to a correlation between the top quark and Higgs masses,
as discussed in Section~\ref{sec:ewk}.  Returning to the CKM matrix itself, we
note in Section~\ref{sec:imp} several ways to obtain improved information on
magnitudes and phases of its elements. Among these is the study of $CP$
violation in the decays of neutral $B$ mesons.  Recent progress in identifying
the flavor of neutral $B$ mesons is reported in Section~\ref{sec:tag}.  We
conclude in Section~\ref{sec:con}. 

\section{Frameworks}\label{sec:fra}

\subsection{The CKM Matrix}

The weak charge-changing interactions lead primarily to the transtions $u
\leftrightarrow d,~c \leftrightarrow s,~t \leftrightarrow b$ between
left-handed quarks ($u,~c,~t$) of charge 2/3 and those ($d,~s,~b$) of charge
$-1/3$.  However, as noted by Cabibbo \cite{NC} and Glashow-Iliopoulos-Maiani
\cite{GIM} for two families of quarks and by Kobayashi and Maskawa
\cite{KM} for three, additional transitions of lesser strength can be
incorporated into this framework in a universal manner.  The charge-changing
transitions then connect $u,~c,~t$ not with $d,~s,~b$ but with a rotated
set $(d',~s',~b') = V (d,~s,~b)$, where $V$ is a unitary $3 \times 3$ matrix
now known as the Cabibbo-Kobayashi-Maskawa (CKM) matrix.

The elements of $V$ are as mysterious as the quark masses, and are intimately
connected with them since the matrix arises as a result of diagonalization of
the quark mass matrices (see, e.g., Ref.~\cite{CKM}).  Moreover, the phases
in the matrix are candidates for the source of $CP$ violation as observed in
the decays of neutral kaons.  We shall assume that to be the case in the
present analysis.

\subsection{Precise electroweak tests}

The top quark plays an indirect role in the extraction of CKM parameters
from data on $B - \bar B$ mixing and $CP$-violating $K \bar K$ mixing.
Thus, it is important to know its mass.  At this time this talk was given,
the best source of information on the top quark was its indirect effects
on the $W$ and $Z$ bosons' self-energies. Some updated information may be found
at the end of Section~\ref{sec:top}. 

In addition to diagrams involving top quarks, $W$ and $Z$ self-energies can be
affected by Higgs bosons and by various new particles which can appear in loop
diagrams.  In conjunction with measurement of the top quark mass, precise
electroweak tests then will be able to shed first light on these contributions.

\section{Determination of CKM parameters}\label{sec:CKM}

We turn now to a description of the CKM matrix elements.  More details
on the measurement of the elements $V_{cb}$ and $V_{ub}$ may be found in
Refs.~\cite{STO,STF}.

\subsection{Parametrization}

We adopt a convention in which quark phases are chosen \cite{BJD} so that the
diagonal elements and the elements just above the diagonal are real and
positive.  The parametrization we shall introduce and employ is one suggested
by Wolfenstein \cite{WOL}. 

The diagonal elements of $V$ are nearly $1$, while
the dominant off-diagonal elements are $V_{us} \simeq -
V_{cd} \simeq \sin \theta \equiv \lambda \simeq 0.22$. Thus to order
$\lambda^2$, the upper $2 \times 2$ submatrix of $V$ is already known from the
Cabibbo-GIM four-quark pattern: 
\begin{equation}
\label{eqn:CGIM}
V \simeq
\left (
\begin{array}{c c c}
1 - \frac{\lambda^2}{2} & \lambda & \cdot \\
- \lambda & 1 - \frac{\lambda^2}{2} &  \cdot \\
\cdot & \cdot & 1
\end{array}
\right )~~~~~~.
\end{equation}
The empirical observation that $V_{cb} \simeq 0.04$ allows one to express it as
$A \lambda^2$, where $A = {\cal O}(1)$. Unitarity then requires $V_{ts} \simeq
- A \lambda^2$ as long as $V_{td}$ and $V_{ub}$ are small enough (which they
are).  Finally, $V_{ub}$ appears to be of order $A \lambda^3 \times {\cal
O}(1)$. Here one must allow for a phase, so one must write $V_{ub} = A
\lambda^3 (\rho - i \eta )$. Finally, unitarity specifies uniquely the form
$V_{td} = A \lambda^3 (1 - \rho - i \eta )$. To summarize, the CKM matrix may
be written 
\begin{equation}
\label{eqn:wolf}
V \approx \left [ \matrix{1 - \lambda^2 /2 & \lambda & A \lambda^3 ( \rho -
i \eta ) \cr
- \lambda & 1 - \lambda^2 /2 & A \lambda^2 \cr
A \lambda^3 ( 1 - \rho - i \eta ) & - A \lambda^2 & 1 \cr } \right ]~~~~~ .
\end{equation}

We shall note below that $V_{cb} = 0.038 \pm 0.005$, so that $A = 0.79 \pm
0.09$. The measurement of semileptonic charmless $B$ decays \cite{ARB,CLB}
gives $|V_{ub}/V_{cb}|$ in the range from 0.05 to 0.11, where most of the
uncertainty is associated with the spread in models \cite{ALT,MOD} for the
lepton spectra. Taking $0.08 \pm 0.03$ for this ratio, we find that the
corresponding constraint on $\rho$ and $\eta$ is $(\rho^2 + \eta^2 )^{1/2} =
0.36 \pm 0.14$. 

The form (\ref{eqn:wolf}) is only correct to order $\lambda^3$ in the matrix
elements. For certain purposes it may be necessary to exhibit corrections of
higher order to the elements.

\begin{figure}
% \vspace{1.5in}
\centerline{\epsfysize = 1.5in \epsffile {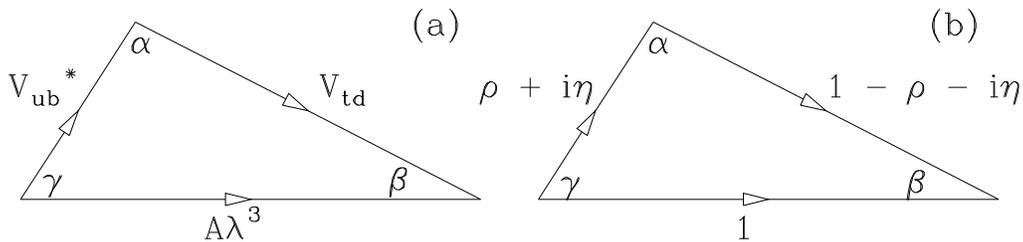}}
\caption{\label{fig:ut} The unitarity triangle.  (a) Relation obeyed by CKM
elements; (b) relation obeyed by (CKM elements)/$A \lambda^3$} 
\end{figure}

The unitarity of $V$ implies that the scalar product of any row and the complex
conjugate of any other row, or of any column and the complex conjugate of any
other column, will be zero.  In particular, taking account of the fact that
$V_{ud}$ and $V_{tb}$ are close to 1, we have
\begin{equation}
V_{ub}^* + V_{td} \simeq A \lambda^3
\end{equation}
or, to the order of interest in small parameters,
\begin{equation}
\rho + i \eta + (1 - \rho - i \eta) = 1~~~.
\end{equation}
The point $(\rho,\eta)$ forms the apex of a triangle in the complex plane,
whose other vertices are the points $(0,0)$ and $(1,0)$.  This
``unitarity triangle'' \cite{UT} and its angles are depicted in
Fig.~\ref{fig:ut}.

The main indeterminacy in the CKM matrix concerns the magnitude of $V_{td}$,
for which only indirect evidence exists.  Correspondingly, we are still quite
uncertain about Arg $V_{ub}^* = \arctan(\eta/\rho)$.  Most of our effort will
be devoted to seeing how these quantities can be pinned down better.

\subsection{Measuring the Cabibbo-GIM submatrix}

The elements of the $2 \times 2$ submatrix connecting the quarks $u,~c$ of
charge 2/3 with those ($d,~s$) of charge $-1/3$ are described satisfactorily by
the single parameter $\lambda$ in the form (\ref{eqn:CGIM}).  The only lingering
question is whether $|V_{ud}|^2 + |V_{us}|^2$ really is 1 (up to corrections of
order $|V_{ub}|^2$, which are negligible), and present data appear to be
consistent with this \cite{CKM,MAR}. 

\subsection{Measuring $V_{cb}$}

In this subsection and the next we give a cartoon version of a discussion which
is set forth much more completely in Ref.~\cite{STF}. 

The decay of a $b$ quark to a charmed quark $c$ and a lepton pair offers the
best hope for determining $V_{cb}$.  One would use the $b$ quark lifetime and
the branching ratio for the process $b \to c \ell \nu$ to estimate the rate for
the process, which would then be proportional to a known kinematic factor times
$|V_{cb}|^2$. Even if $b$ and $c$ were free, we would have to know their masses
accurately in order to make a useful estimate.

Since the $b$ quark and charmed quark are incorporated into hadrons such as a
$B$ meson and a $D$ meson, the problem becomes one of estimating hadronic
effects.  There are several ways to do this.

\subsubsection{Free quarks}

A good deal of indeterminacy of the rate for $b \to c \ell \bar \nu_\ell$ is
associated merely with uncertainty in quark masses.  However, the uncertainty
in the predicted decay rate can be reduced by constraints on the mass
difference $m_b - m_c$ from hadron spectroscopy \cite{KIM,TASI}.  Taking $m_b
= 5.0 \pm 0.3$ GeV/$c^2$, $m_b - m_c$ ranging from 3.34 to 3.40 GeV/$c^2$, $B(B
\to {\rm charm} + \ell + \bar \nu_l) = 10.5\%$, and $\tau_B = 1.49$ ps, we
obtained \cite{CKM} $V_{cb} = 0.038 \pm 0.003$. 

\subsubsection{Free quarks and QCD}

One can take account of the effects of the light quarks by means of Fermi
momentum and can apply QCD corrections to the decay of the free $b$ quark
\cite{ALT}. The result should be an average over the excitation of individual
final states of the charmed quark and the spectator antiquark. 

\subsubsection{Models for final states}

One can calculate $B$ semileptonic decay rates to specific final states, such
as $D \ell \bar \nu_\ell,~D^* \ell \bar \nu_\ell$, and so on \cite{MOD}. It
is then necessary to include all relevant states, so an important question is
what charmed states besides $D$ and $D^*$ play a role.

\subsubsection{Use of the ``zero-recoil'' point}

When the lepton pair has its maximum invariant mass, the $b$ quark decays to a
charmed quark without causing it to recoil \cite{NOR}.  Thus, hadronic
effects are kept to a minimum.  The limitation on this method is mainly one of
statistics at present. 

\subsubsection{Averages}

When the various methods are combined, one gets an idea of the spread in
theoretical approaches.  In Ref.~\cite{CKM} we quoted the value $V_{cb} =
0.038 \pm 0.005$ obtained in Ref.~\cite{STO} on the basis of such averages. 
That is the value which we will use in the present analysis, corresponding to
$A = 0.79 \pm 0.09$.  More recently Stone \cite{STF} estimates $V_{cb} =
0.038 \pm 0.003$, in accord with our original free-quark value and
corresponding to an error $\Delta A = 0.06$. 

\subsection{Measuring $V_{ub}$}

In order to see the effects of the process $b \to u \ell \bar \nu_\ell$,
one has to study leptons beyond the end point for charm production.  As
a result, one sees only a very small part of the total phase space for
the process of interest.  The question then becomes one of how the decay
populates this small region of phase space.  The final $u$ quark can
combine with the initial $\bar u$ or $\bar d$ in the decaying $B$ meson to
form a nonstrange hadron such as $\pi,~\rho,~a_1,~\ldots$.  One can either
describe this recombination in an average sense \cite{ALT,DON} or
employ models for excitation of individual resonances \cite{MOD}.

The range of theoretical approaches allows values of $|V_{ub}/V_{cb}|$
between 0.05 and about 0.11 when the more recent CLEO data are used
\cite{CLB}. Somewhat larger values (up to a factor of 2, in some models) are
implied by earlier ARGUS data \cite{ARB}.  These results correspond to a
fraction of $b$ decays without charmed particles of between 1 and 2\%.  For
present purposes we shall take $|V_{ub}/V_{cb}| = 0.08 \pm 0.03$.  For
comparison, Stone \cite{STF} quotes $|V_{ub}/V_{cb}| = 0.08 \pm 0.02$. 

\subsection{Arg $V_{ub}$}

The phase of $V_{ub}$ is one of the least well known parameters of the CKM
matrix.  For it, we must rely upon indirect information.

\subsubsection{$B^0 - \overline{B}^0$ mixing}

\begin{figure}
% \vspace{1.3in}
\centerline{\epsfysize = 1.3in \epsffile {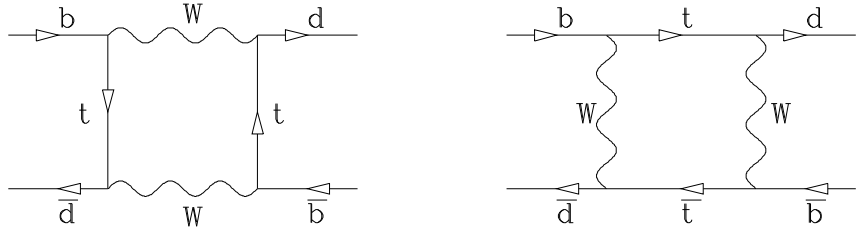}}
\caption{\label{fig:bbox} Dominant box diagrams for mixing of $B^0$ and $\bar
B^0$} 
\end{figure}

The original evidence for $B^0 - \bar B^0$ mixing came from the observation
\cite{ARM} of ``wrong-sign'' leptons in $B$ meson semileptonic decays.
The diagrams of Fig.~\ref{fig:bbox} give rise to a splitting between mass
eigenstates
\begin{equation} 
\label{eqn:bmix}
\Delta m \sim f_B^2 m_t^2 |V_{td}|^2
\end{equation}
times a slowly varying function of $m_t$.  (See, e.g., Ref.~\cite{CKM} for
detailed expressions.)  Here $f_B$ is the ``$B$ meson decay constant,'' which
expresses the overlap of a $b \bar q$ state at zero separation with the wave
function of the $B$ meson.  Information on $f_B$ is improving, but still is a
major source of indeterminacy.  The top quark mass $m_t$ is becoming better
known, as we shall see at the end of Section~\ref{sec:top}.  The CKM element
$V_{td}$ has a magnitude proportional to $|1 - \rho - i \eta|$, which is what
we would like to learn. 

\begin{figure}
% \vspace{1.75in}
\centerline{\epsfysize = 1.75in \epsffile {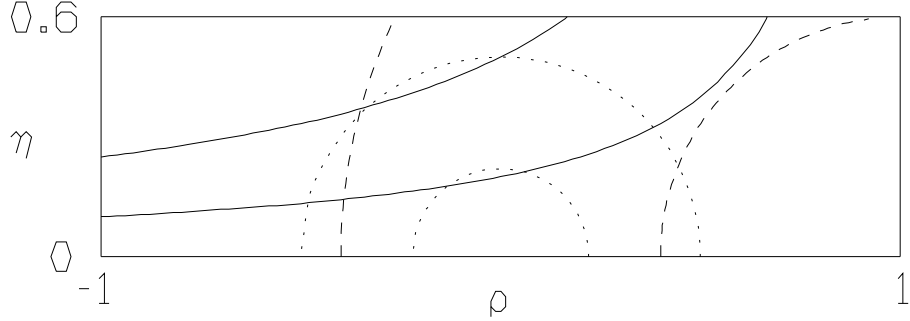}}
\caption{Shapes of allowed regions in $(\rho,\eta)$ plane arising from $B -
\bar B$ mixing (dashes), $CP$-violating $K - \bar K$ mixing (solid), and
$|V_{ub}/V_{cb}|$ (dots)} 
\label{fig:region}
\end{figure}

The average value of data used for the present analysis gives $\Delta m /
\Gamma = 0.66 \pm 0.10$ \cite{MXA}.  The resulting constraint on $(\rho,
\eta)$ for fixed values of $f_B$ and $m_t$ is a circular band with radius
approximately 1 and center at the point (1,0). Uncertainty in $f_B$ and, to a
lesser extent, $m_t$, is a source of spread in this band, whose shape is
illustrated by the dashed arcs in Fig.~\ref{fig:region}. 

\begin{figure}
% \vspace{1.3in}
\centerline{\epsfysize = 1.3in \epsffile {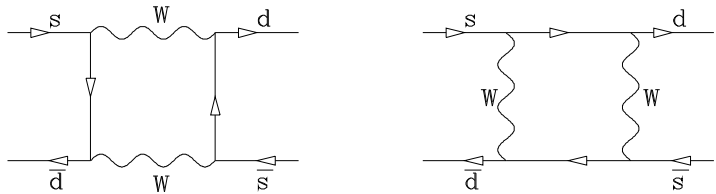}}
\caption{Box diagrams for mixing of $K^0$ and $\bar K^0$}
\label{fig:kbox}
\end{figure}

\subsubsection{$CP$-violating $K^0 - \overline{K}^0$ mixing}

The box diagrams shown in Fig.~\ref{fig:kbox} give rise to a $CP$-violating
term in the matrix element between a $K^0$ and a $\bar K^0$, 
\begin{equation}
\label{eqn:kmix}
{\rm Im}~{\cal M}_{12} \sim f_K^2~{\rm Im}~(V_{td}^2) \sim \eta (1-\rho)~~~,
\end{equation}
so that the constraint in the $(\rho,\eta)$ plane is a band bounded by
hyperbolae with focus at the point (1,0), as illustrated by the the example of
the solid lines in Fig.~\ref{fig:region}.  Here, again, the top quark mass
enters.  There are small corrections (not completely negligible) from charmed
quarks in the loop. Neglecting these, however, one can take the quotient of the
constraints (\ref{eqn:bmix}) and (\ref{eqn:kmix}) to find a constraint on Arg
$V_{td}$.  As we shall see, such a constraint is useful in predicting the
expected asymmetry in certain $CP$-violating decays of $B$ mesons. 

\subsection{Allowed region of parameters}

\begin{figure}
% \vspace{3.4in}
\centerline{\epsfysize = 3.4in \epsffile {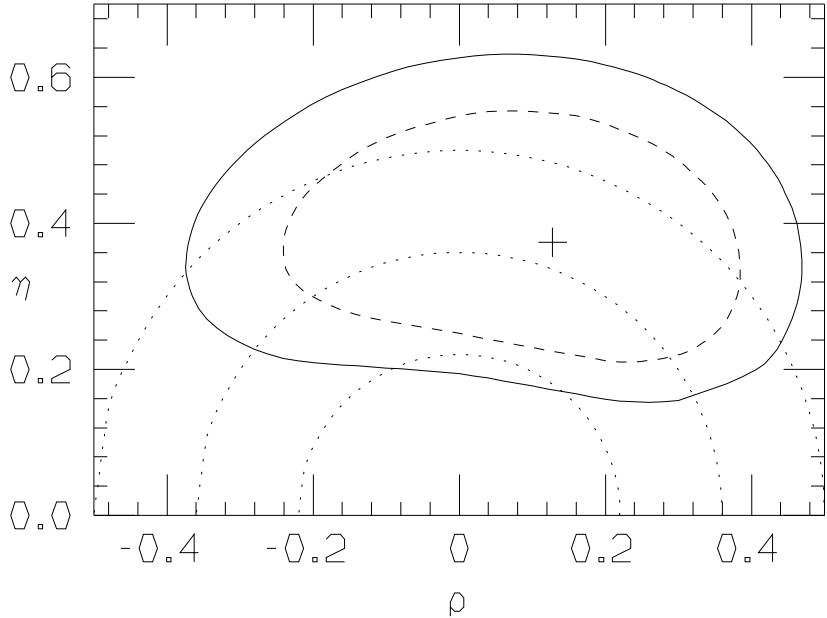}}
\caption{\label{fig:potato} Contours of 68\% (inner curve) and 90\% (outer
curve) confidence levels for regions in the $(\rho,\eta)$ plane.  Dotted
semicircles denote central value and $\pm 1 \sigma$ limits implied by
$|V_{ub}/V_{cb}| = 0.08 \pm 0.03$.  Plotted point corresponds to minimum
$\chi^2 = 0.17$, while (dashed, solid) curves correspond to $\Delta \chi^2 =
(2.3,~4.6)$} 
\end{figure}

When the constraints of Eqs.~(\ref{eqn:bmix}) and (\ref{eqn:kmix}) are combined
with that on $|V_{ub}/V_{cb}|$ [shown as the circular band bounded by the
dotted arcs in Fig.~\ref{fig:region}], one gets the allowed region of
parameters shown in Fig.~\ref{fig:potato} and described by the first line in
Table~\ref{tab:changes}. 

The parameters taken for the present analysis are those chosen in
Ref.~\cite{CKM}, and include the choices $m_t = 160 \pm 30$ GeV, $f_B = 180
\pm 30$ MeV, $|V_{ub}/V_{cb}| = 0.08 \pm 0.03$, and $A = 0.785 \pm 0.093$.  The
allowed region at 90\% c.l. has $-0.4 \le \rho \le 0.5$ for $\eta \simeq 0.3$,
while for $\rho \simeq 0$ one has $0.2 \le \eta \le 0.6$.  For a broad range of
parameters, CKM phases can describe $CP$ violation in the kaon system.  The
question is whether this explanation of the observed $CP$ violation is the
correct one. A partial answer may be obtained by acquiring improved information
about the top quark mass or about CKM elements. (See also
Refs.~\cite{HR,BUR}.) 

The choice of $m_t$ mentioned above was based on an analysis of electroweak
data parallel to that presented in Ref.~\cite{SWA} and reaching the same
conclusions.  We shall give more details in Section~\ref{sec:ewk}. The results
of fixing the top quark mass at 160 or 190 GeV are shown in
Table~\ref{tab:changes}.  The allowed region is shrunk only slightly, and there
is not much difference between the two cases.  The favored value of $\rho$
increases by 0.03 for each 10 GeV increase in $m_t$.

\begin{table}
\begin{center}
\caption{\label{tab:changes}Effects of changing parameters in fits to CKM
matrix elements from ``nominal'' values described in text.  In all fits there
is one degree of freedom} 
\medskip
\begin{tabular}{c c c c c} \hline
Parameters     & $\rho$ & $\eta$ & $\chi^2_{\rm min}$ & $\rho$ range $^{a)}$ \\
   & ($\chi^2_{\rm min}$) & ($\chi^2_{\rm min}$) &      &             \\ \hline
``Nominal''     & 0.13  & 0.37     &      0.17          & --0.37 to 0.47 \\
$m_t = 160$ GeV & 0.13  & 0.37     &      0.17          & --0.31 to 0.45 \\
$m_t = 190$ GeV & 0.22  & 0.34     &      0.25          & --0.19 to 0.50 \\
$B_K = 0.80 \pm 0.02^{~b)}$ & 0.13 & 0.38 &   0.23      & --0.28 to 0.45 \\
$V_{cb} = 0.038 \pm 0.002$ & 0.11 & 0.38 & 0.20         & --0.37 to 0.45 \\
$f_B = 180 \pm 10$ MeV & 0.15 & 0.37 &    0.22      & --0.28 to 0.43 \\ \hline
\end{tabular}
\end{center}
\leftline{$^{a)}$ bounded by solid lines in \protect Fig.~\ref{fig:potato}~~~~
$^{b)}~m_t$ fixed at 160 GeV}
\end{table}

A parameter known as $B_K$ describes the degree to which the diagrams of
Fig.~\ref{fig:kbox} actually dominate the $CP$-violating $K^0 - \bar K^0$
mixing.  In the fits described so far we took the nominal value of $B_K = 0.8
\pm 0.2$.  If we take $m_t = 160$ GeV/$c^2$ and reduce the error on $B_K$ to
0.02, we obtain the result shown in Table~\ref{tab:changes}.  The reduced
errors on $B_K$ are clearly not of much help.  The major errors remaining are
those of $f_B$, $V_{cb}$, and $|V_{ub}/V_{cb}|$. 

We next tried reducing the error on $V_{cb}$ to 0.002, keeping other parameters
as in the original fit.  Again, there is little shrinkage of the allowed
parameter space.  Reduction of the error on $f_B$ from 30 to 10 MeV helps a
little.  The results of these exercises are shown in Table~\ref{tab:changes}.
The shapes of the allowed regions change very little. The conclusion is that
one needs simultaneous reduction in the errors of several observables to
significantly narrow down the range of CKM parameters.  We explore these
possibilities in Section~\ref{sec:imp}. First, however, we concentrate on the
top quark. 

\section{The top quark}\label{sec:top} 

\subsection{Indirect evidence}

Indirect evidence for the top quark has been around for a long time.
The neutral-current couplings of the $b$ quark (both flavor-conserving
and the absence of flavor-violating ones) have persuaded us that the
left-handed $b$ quark is a member of a doublet $(t,b)_L$ of weak SU(2), while
the right-handed $b$ is a singlet of weak SU(2) \cite{LHB}.

The expectation that the top quark is relatively heavy is more recent. A value
of $m_t$ of at least 70 GeV was needed in order to understand the unexpectedly
large magnitude of $B^0 - \bar B^0$ mixing \cite{ARM}. Even a higher lower
bound is required to understand the size of $CP$-violating $K^0 - \bar K^0$
mixing \cite{BKM}.  The branching ratio of the $W$ to (lepton) + (neutrino)
of about 1/9 is compatible with there being no contribution from $W \to t +
\bar b$, indicating that $m_t > M_W - m_b$.  

An upper limit on the top quark mass is provided by its effects on $W$ and
$Z$ self-energies.  In the lowest-order electroweak theory, a measurement
of the $Z$ mass implies a specific value of $M_W$.  The $W$ and $Z$
self-energies are affected by top quark and Higgs masses, so that now
$M_W/M_Z = f(m_t,~m_{\rm Higgs})$.  This function is quadratic in $m_t$ but
only logarithmic in $M_{\rm Higgs}$.  When the Higgs boson mass is allowed to
range up to 1 TeV (above which the theory should dynamically generate a mass of
1 -- 2 TeV in any case), the observed values of $M_W$ and of many other
electroweak observables allow one to conclude that $m_t \le 200$ GeV/$c^2$. 

\subsection{Direct searches}

The signature for top quark pair production in $\bar p p$ collisions is
the simultaneous decay $t \to W^+ + b,~\bar t \to W^- + \bar b$.  One
channel with little background involves the decay of one $W$ to $e \nu$ and the
other to $\mu \nu$.  As of this workshop, the CDF Collaboration had identified
two $e \mu$ candidates and the D0 Collaboration had observed one.  On this
basis, all that were quoted were lower limits on the top quark mass.  Using a
sample in which hadronic decays of one of the two $W$'s were also searched
for, D0 quoted a lower limit \cite{DTL} of 131 GeV/$c^2$.

\subsection{Postscript: evidence}

Since this workshop, the CDF Collaboration has presented evidence for
the production of a top quark \cite{CDFTE} with $m_t = 174 \pm 10
~~^{+13}_{-12}$ GeV/$c^2$.  The cases we chose of $m_t = 160$ and $190$
GeV/$c^2$ are compatible with this value.  The main impact of this measurement
is felt less on the determination of CKM parameters than on the interpretation
of electroweak results, which we discuss next. 

\section{Impact of electroweak tests}\label{sec:ewk}

\begin{figure}
% \vspace{3.4in}
\centerline{\epsfysize = 3.4in \epsffile {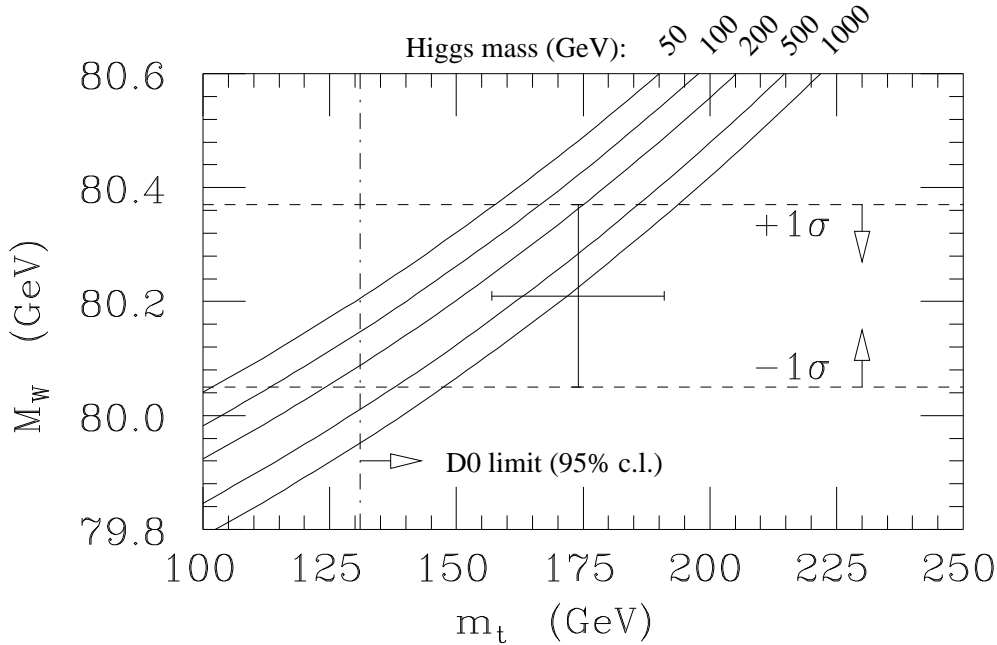}}
\caption{\label{fig:mwmt}Dependence of $W$ boson mass $M_W$ on top quark
mass $m_t$ for various values of Higgs boson mass (labels on curves).
[Postscript:  The plotted point denotes the world average \protect \cite{MWA}
of direct $W$ mass measurements, $M_W = 80.23 \pm 0.18$ GeV/$c^2$, and the
recent CDF top mass value \protect \cite{CDFTE} of $m_t = 174 \pm 17$
GeV/$c^2$.]}
\end{figure}

In Fig.~\ref{fig:mwmt} we show the electroweak prediction for $M_W$ as a
function of $m_t$ for various values of Higgs boson mass $M_H$.  Also shown is
the latest $1 \sigma$ range of $M_W$, corresponding to the average over many
experiments \cite{MWA}. (The latest results have been presented by the CDF
and D0 collaborations.)  Even without a direct observation, one sees the 
upper bound of about 200 GeV/$c^2$ quite clearly.  The recent (post-workshop)
observation corresponds to a data point lying in the allowed range. Greater
precision on both $M_W$ and $m_t$ will be needed to distinguish among
possibilities for Higgs boson masses. 

\begin{figure}
% \vspace{3.1in}
\centerline{\epsfysize = 3.1in \epsffile {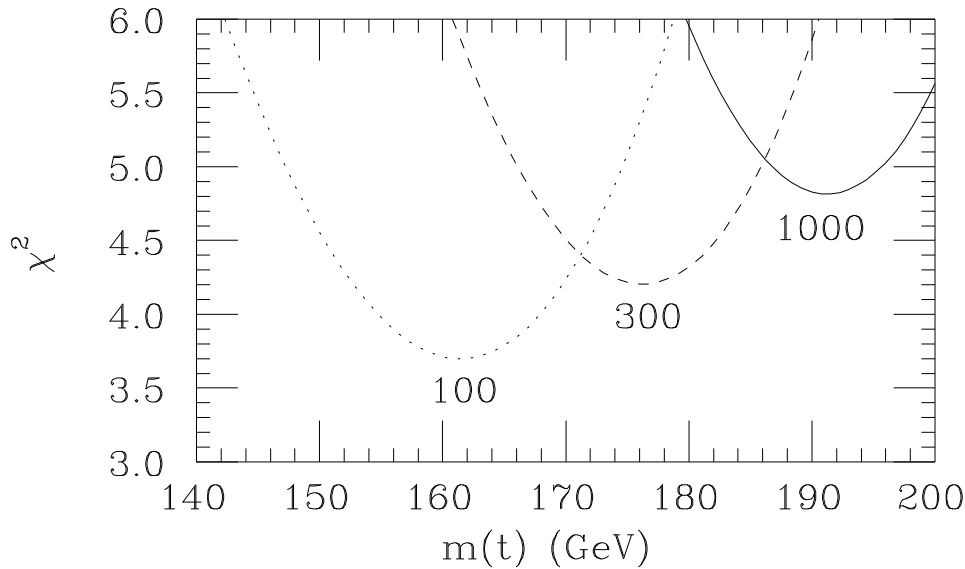}}
\caption{\label{fig:chi} Behavior of $\chi^2$ as function of top quark mass
for Higgs boson masses of 100, 300, and 1000 GeV (labels on curves)}
\end{figure}

A fit to the electroweak observables cited in Table~\ref{tab:obs} has been
performed. In each case a prediction is made for the ``nominal'' values of $m_t
= 140$ GeV and $M_H = 100$ GeV/$c^2$.  The Higgs boson mass is held fixed,
while the top quark mass is allowed to vary in such a way as to minimize the
$\chi^2$ of the fit.  The results are shown in Fig.~\ref{fig:chi} and
Table~\ref{tab:val}.

\begin{table}
\begin{center}
\caption{\label{tab:obs}Electroweak observables described in fit}
\medskip
\begin{tabular}{c c c c} \hline
Quantity        &   Experimental   &   Nominal    &  Experiment/     \\
                &      value       &    value     &   Nominal        \\ \hline
$Q_W$ (Cs)      & $-71.0 \pm 1.8^{~a)} $  &   $ -73.2^{~b)}$
   & $0.970 \pm 0.025$ \\
$M_W$ (GeV/$c^2$) & $80.22 \pm 0.14^{~c)}$  & $80.174^{~d)}$
   & $1.001 \pm 0.002$ \\
$\Gamma_{\ell\ell}(Z)$ (MeV) & $83.82 \pm 0.27^{~e)}$ & $83.6^{~f)}$
   & $1.003 \pm 0.003$ \\
$\Gamma_{\rm tot}(Z)$ (MeV) & $2489 \pm 7^{~e)}$ & $2488 \pm 6^{~f)}$
   & $1.000 \pm 0.004$\\
$\sin^2 \hat \theta_W^{\rm eff}$ & $0.2318 \pm 0.0008^{~g)}$ & $0.2322^{~f)}$
   & $0.998 \pm 0.003$ \\
$\sin^2 \hat \theta_W^{\rm eff}$ & $0.232 \pm 0.009^{~h)}$ & 0.2322 &
   $0.999 \pm 0.039$ \\
$\sin^2 \hat \theta_W^{\rm eff}$ & $0.2287 \pm 0.0010^{~i)}$ & 0.2322 &
   $0.985 \pm 0.004$ \\ \hline
\end{tabular}
\end{center}
\leftline{$^{a)}$ Weak charge in cesium.  From \protect Ref.~\cite{QWE}}
\leftline{$^{b)}$ From \protect Ref.~\cite{MR}, incorporating corrections of
\protect Ref.~\cite{QWT}}
\leftline{$^{c)}$ Average of direct measurements from \protect Ref.~\cite{MWA}
and indirect information}
\leftline{\quad from neutral/charged current ratio in
deep inelastic neutrino scattering \protect \cite{NCR}}
\leftline{$^{d)}$ As calculated in \protect Ref.~\cite{DKS}}
\leftline{$^{e)}$ LEP average as of August, 1993 \protect \cite{SWA}}
\leftline{$^{f)}$ As calculated in \protect Ref.~\cite{MR}}
\leftline{$^{g)}$ From asymmetries at LEP, containing corrections of
\protect Ref.~\cite{GS}}
\leftline{$^{h)}$ From $\nu_\mu e$ and $\bar \nu_\mu e$ scattering
\protect \cite{CHA}}
\leftline{$^{i)}$ From left-right asymmetry in annihilations at SLC
\protect \cite{ALR}, containing}
\leftline{\quad corrections of \protect Ref.~\cite{GS}}
\end{table}

\begin{table}
\begin{center}
\caption{\label{tab:val}Values of $m_t$ for $\chi^2$ minima in fits to
electroweak observables} 
\medskip
\begin{tabular}{c c c} \hline
$M_H$ (GeV/$c^2$) & $m_t$ (GeV/$c^2$) & $\chi^2$ \\ \hline
        100       &   $161 \pm 12$    &     3.7  \\
        300       &   $176 \pm 11$    &     4.2  \\
       1000       &   $191 \pm 10$    &     4.8  \\ \hline
\end{tabular}
\end{center}
\end{table}

A slight preference is shown for a light Higgs boson.  This is driven
in part by the low value of $\sin^2 \hat \theta_W^{\rm eff}$ obtained at
SLC/SLD.

The range of top quark masses we chose to discuss at the workshop is compatible
both with the results of Table~\ref{tab:val} and with the recently announced
observation.  Present errors on $m_t$ do not allow a conclusion to be drawn yet
about the Higgs boson mass. 

\section{Improved CKM Information}\label{sec:imp}

\subsection{Meson decay constants}

As we mentioned earlier, uncertainty in $f_B$ is a major source of
indeterminacy in extracting $|V_{td}|$ from $B^0 - \bar B^0$ mixing.  Early
compilations of predictions are contained in Refs.~\cite{FBR}.  More recent
information on meson decay constants has been provided by lattice gauge theory
\cite{FBL}, QCD sum rules \cite{FBQ}, and direct quark-model calculations
\cite{FBM} which make use of spin-dependent electromagnetic mass splittings
in charmed and $B$ mesons to estimate the wave function of a light-heavy system
at zero interquark separation.  One can expect the reliability of the lattice 
and QCD sum rule calculations to improve as they are tested on a wide range
of properties of charmed and $b$-flavored hadrons, while the quark model
estimate would be helped by a precision measurement of isospin splittings
in $B$ and $B^*$ mesons.  Modest improvements of recent measurements of the
decay constant $f_{D_s}$ \cite{FDS} will allow one to check these schemes.

\subsection{Rare kaon decays}

Several rare kaon decays can provide information on CKM parameters.  Here we
discuss the decays of kaons to a pion and a lepton pair.

The rate for the process $K^+ \to \pi^+ \nu \bar \nu$ (summed over neutrino
species) is sensitive to $|V_{td}|^2$.  Very roughly, for a nominal top quark
mass of 140 GeV/$c^2$, a branching ratio of less than $10^{-10}$ favors $\rho >
0$ while a branching ratio of greater than $10^{-10}$ favors $\rho < 0$.  The
branching ratio is an increasing function of top quark mass.  Details have
been given in Refs.~\cite{CKM,HR}.

The present experimental limit \cite{KPN}, $B(K^+ \to \pi \nu \bar \nu) <
5 \times 10^{-9}~(90\%$ c.l.) is a factor of 50 above the expected level,
but further improvements in data collection are foreseen.

[Postscript:  Here information on $m_t$ is very welcome; the predictions were
quoted at the workshop for $m_t = 100,~140$, and 180 GeV/$c^2$. An updated set
of predictions may be found in Ref.~\cite{BUR}.  It now seems more likely
that the branching ratio will be at least $10^{-10}$.]

The decays $K_L \to \pi^0 \ell^+ \ell^-$ are expected to proceed mainly through
$CP$ violation \cite{GL}, while key $CP$-conserving backgrounds to this
process (see, e.g., Ref.~\cite{HG}) are absent in $K_L \to ~\pi^0 \nu \bar
\nu$.  Present 90\% c.l. upper limits on the branching ratios for these
processes are shown in Table~\ref{tab:brs}.  The expected branching ratios
\cite{DIB} are about $10^{-11}$. 

\begin{table}
\begin{center}
\caption{\label{tab:brs}Upper limits on branching ratios for decays of neutral
kaons to neutral pions and a lepton pair} 
\medskip
\begin{tabular}{c c c} \hline
Process & 90\% c.l.   &  Reference \\
        & upper limit &            \\ \hline
$K_L \to \pi^0 e^+ e^-$ & $1.8 \times 10^{-9}$ & \cite{E845,DHE,E731} \\
$K_L \to \pi^0 \mu^+ \mu^-$ & $5.1 \times 10^{-9}$ & \cite{DHM} \\
$K_L \to \pi^0 \nu \bar \nu$ & $2.2 \times 10^{-4}$ & \cite{GG} \\ \hline
\end{tabular}
\end{center}
\end{table}

\subsection{$CP$ violation in decays of neutral kaons}

One can search for a difference between the $CP$-violation parameters
$\eta_{+-} = \epsilon + \epsilon'$ and $\eta_{00} = \epsilon - 2 \epsilon'$ in
the decays of neutral kaons to pairs of charged and neutral pions.  A non-zero
value of $\epsilon'/\epsilon$ would confirm predictions of the CKM origin of
$CP$ violation in the kaon system, and has long been viewed as one of the most
promising ways to disprove a ``superweak'' theory of this effect
\cite{SWK,RVW}. 

The latest estimates by A. Buras and collaborators \cite{BEPS} are equivalent
to $[\epsilon'/\epsilon]|_{\rm kaons} = (1/2~~{\rm to}~~3) \times 10^{-3}
\eta$, with smaller values for higher top quark masses. The Fermilab E731
Collaboration \cite{GIB} measures $\epsilon'/\epsilon = (7.4 \pm 6) \times
10^{-4}$, leading to no restrictions on $\eta$ in comparison with the range
(0.2 to 0.6) we have already specified. The CERN NA31 Collaboration
\cite{NA31} finds $\epsilon'/\epsilon = (23 \pm 7) \times 10^{-4}$,
consistent only with $\eta \stackrel{>}{\sim} 1/2$ and a light top quark.
[Postscript:  this scenario now appears less likely in view of the result of
Ref.~\cite{CDFTE}.]  Both groups are preparing new experiments, for which
results should be available around 1996. 

\subsection{Rare $B$ decays}

The rate for the purely leptonic process $B \to \ell \bar \nu_\ell$ provides
information on the combination $f_B |V_{ub}|$.  One expects a branching ratio
of about $10^{-4}$ for $\tau \bar \nu_\tau$ and $(1/2) \times 10^{-6}$ for $\mu
\bar \nu_\mu$.  A suggestion was made \cite{PFH} for eliminating $f_B$ by
comparing the $B \to \ell \bar \nu_\ell$ rate with the $B^0 - \bar B^0$ mixing
amplitude, and thereby measuring the ratio $|V_{ub}/V_{td}|$ directly.  While
such a measurement is unlikely to tell whether the unitarity triangle has
nonzero area (and thus whether the CKM phase is the origin of $CP$ violation in
the kaon system), it {\it can} help resolve ambiguity regarding the value of
$\rho$. 

Another interesting ratio \cite{ALI} is the quantity $\Gamma(B \to \rho
\gamma)/ \Gamma(B \to K^* \gamma)$, which, aside from small phase space
corrections, should just be $|V_{td}/V_{ts}|^2 \simeq 1/20$. 

\subsection{$B_s - \bar B_s$ mixing}

The mixing of $B_s$ and $\bar B_s$ via diagrams similar to those in
Fig.~\ref{fig:bbox} involves the combination $f_{B_s}^2|V_{ts}|^2$ instead of
$f_B^2|V_{td}|^2$. Since we expect $|V_{ts}| \approx |V_{cb}| \approx 0.04$,
the main uncertainties in $x_s \equiv (\Delta m/\Gamma)|_{B_s}$ are associated
with $f_{B_s}$ and $m_t$.  A range of 10 to 50 is possible for this quantity.
Alternatively, one can estimate the ratio $f_{B_s}/f_B$ using models for SU(3)
symmetry breaking, and one finds \cite{CKM} $x_s = (19 \pm 4)/[(1 - \rho)^2 +
\eta^2]$.  With the values of $\rho$ and $\eta$ suggested by present fits to
data, the most likely value of $x_s$ seems to be around 20.  This corresponds
to many oscillations between $B_s$ and $\bar B_s$ over the course of a $B_s$
lifetime (about 1.5 ps), and represents a strong experimental challenge. 

\subsection{$CP$ violation in $B$ systems}

Asymmetries in the rates for certain decays of $B$ mesons can provide direct
information about the angles in the unitarity triangle of Fig.~\ref{fig:ut}. 
These decays involve final states which are eigenstates of $CP$, so that they
can be reached both from an initial $B^0$ and from an initial $\bar B^0$. 

\begin{figure}
% \vspace{3in}
\centerline{\epsfysize = 3in \epsffile {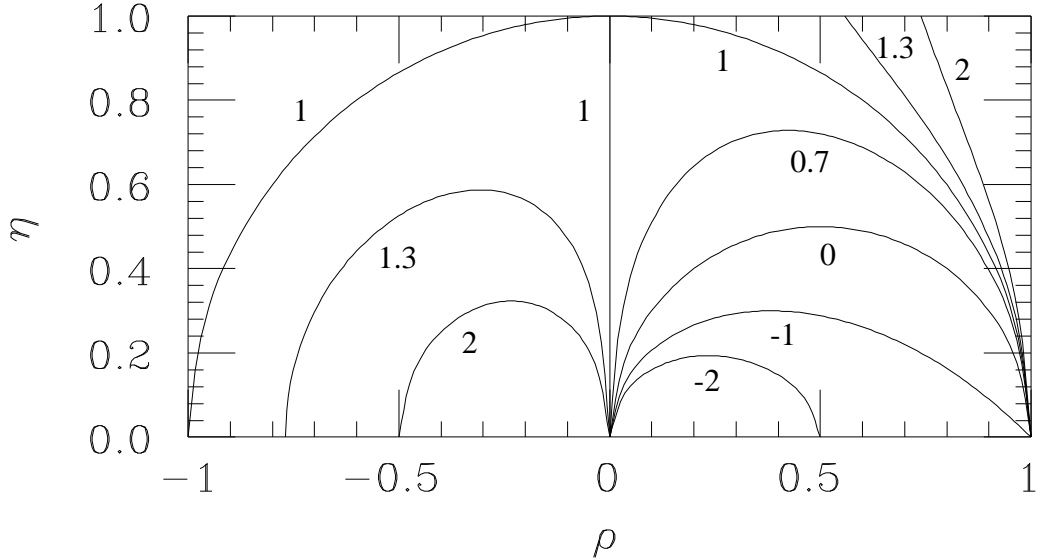}}
\caption{\label{fig:rat} Contours of ratios of asymmetries ${\cal R} \equiv 
A(\pi^+ \pi^-)/ A(J/\psi K_S) = \sin (2 \alpha) / \sin(2 \beta)$ (labels on
curves) in the $(\rho,\eta)$ plane} 
\end{figure}

We may define a time-integrated rate asymmetry $A(f)$ as
\begin{equation}
A(f) \equiv \frac{\Gamma(B^0_{t=0} \to f) - \Gamma(\bar B^0_{t=0} \to f)}
{\Gamma(B^0_{t=0} \to f) + \Gamma(\bar B^0_{t=0} \to f)}~~~.
\end{equation}
The angles $\beta$ and $\alpha$ in Fig.~1 are related to the asymmetries
in decays to $J/\psi K_S$ and $\pi^+ \pi^-$ final states:
\begin{equation}
A(J/\psi K_S) = - \frac{x_d}{1 + x_d^2} \sin 2 \beta~~~,
\end{equation}
\begin{equation}
A(\pi^+ \pi^-) = - \frac{x_d}{1 + x_d^2} \sin 2 \alpha~~~,
\end{equation}
where $x_d \equiv (\Delta m/\Gamma)|_{B^0}$, and we have neglected lifetime
differences between eigenstates.  Contours of the expected values of these
asymmetries have been quoted in Refs.~\cite{CKM,HR} and will not be
reproduced here.  The {\it ratios} of these asymmetries can be useful in
cancelling certain common (and sometimes hard-to-estimate) ``dilution
factors'' associated with identification of the flavor of the decaying
neutral $B$ meson.  Contours \cite{PFH} of the ratio ${\cal R} \equiv \sin 2
\alpha / \sin 2 \beta$ are shown in Fig.~\ref{fig:rat}.  As long as $\rho^2 +
\eta^2 < 1$, which certainly is true, a value of ${\cal R} > 1$ signifies $\rho
< 0$, while ${\cal R} < 1$ signifies $\rho > 0$. 

\section{Recent results on tagging neutral $B$ mesons}\label{sec:tag}

\subsection{Why neutral $B$ mesons?}

The observation of a $CP$-violating asymmetry between the rate for a process
and its charge-conjugate requires some sort of interference.  Two examples
serve to illustrate the major possibilities \cite{BCP}.

\subsubsection{Self-tagging modes}

The rates for such processes as $B^+ \to K^+ \pi^0$ and $B^- \to K^- \pi^0$ can
differ from one another.  Under charge-conjugation, weak phases change sign
but strong phases do not.  One can see a $CP$-violating rate difference, but
only if strong phases differ in the $I = 1/2$ and $I = 3/2$ channels.
Interpretation of an effect requires knowledge of this final-state phase
difference.  [Postscript:  with the help of SU(3) symmetry and some simplifying
assumptions, it is possible to extract CKM phases from rates of self-tagging
modes alone \cite{BPP}.]

\subsubsection{Decays to a $CP$ eigenstate}

Final-state phase information is not needed if one compares the rates for
a state which is produced as a $B^0$ and a state which is produced as a $\bar
B^0$ to decay to an eigenstate $f$ of $CP$.  The relevant interference leading
to a rate asymmetry occurs between amplitudes for decay and $B^0 - \bar B^0$
mixing.  As mentioned above, decay rate asymmetries can directly probe the
angles of the unitarity triangle, as long as a single amplitude contributes to
each transition $B^0 \to f$ and $\bar B^0 \to f$.  To make use of this method,
one must identify the flavor of the decaying particle at the time of
production:  was it a $B^0$ or a $\bar B^0$?

\subsection{Identifying neutral $B$'s}

\subsubsection{At the $\Upsilon({\bf 4S})$ resonance}

A peak in the cross section for $e^+ e^- \to B^0 \bar B^0$ occurs just above
threshold at the $\Upsilon(4S)$ resonance.  If one ``tags'' the flavor of the
decaying state by observing the semileptonic decay of the ``other'' $B$, the
existence of $B^0 - \bar B^0$ mixing and the correlation of the $B^0$ and $\bar
B^0$ in a state of negative charge-conjugation lead to an asymmetry
proportional to $\sin(t_1 - t_2)$, where $t_1$ is the proper time of the decay
to the $CP$ eigenstate and $t_2$ is the proper time of the tagging decay. This
asymmetry vanishes when integrated over all decay times, so one needs
information on $t_1 - t_2$ such as might be provided by an asymmetric $B$
factory. 

\subsubsection{Away from the $\Upsilon({\bf 4S})$ resonance}

In any reaction in which a $b \bar b$ pair is produced at high energy, such
as a hadronic collision or the decay of the $Z^0$, the flavor of a neutral $B$
meson decaying to a $CP$ eigenstate can be tagged by looking at the flavor of
the $b$-flavored particle produced in association with it.  Such a particle
might be another neutral nonstrange or strange $B$ (in which case mixing would
cause a dilution of tagging efficiency), or it could be a charged $B$ or
a $b$-flavored baryon.  Here one has to find the tagging particle among the
debris of the collision, and estimates of tagging efficiency are likely to
be model-dependent.

\subsubsection{Tagging using associated hadrons}

A neutral $B$ meson produced in a high-energy collision is likely to be
accompanied by other hadrons as a result of the fragmentation of the initial
quark or as a result of cascades from higher resonances. This feature could be
useful for tagging the flavor of a produced $B$ meson \cite{GNR,GRL,GRD}. 

\begin{figure}
% \vspace{1.5in}
\centerline{\epsfysize = 1.5in \epsffile {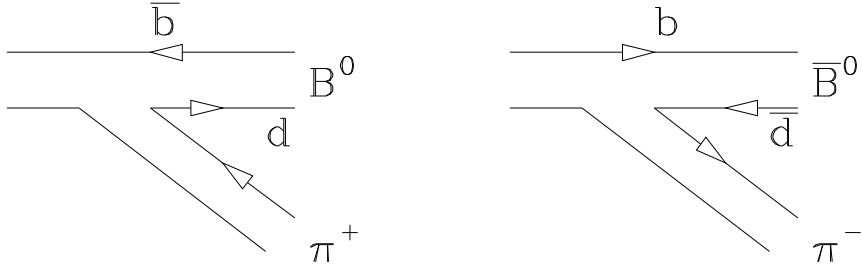}}
\caption{\label{fig:cor} Quark graphs describing correlation between flavor
of neutral $B$ meson and charge of leading pion in fragmentation}
\end{figure}

The correlation is easily visualized with the help of the quark diagrams shown
in Fig.~\ref{fig:cor}.  By convention (the same as for kaons), a neutral $B$
meson containing an initially produced $\bar b$ is a $B^0$.  It also contains a
$d$ quark.  The next charged pion down the fragmentation chain must contain a
$\bar d$, and hence must be a $\pi^+$.  Similarly, a $\bar B^0$ will be
correlated with a $\pi^-$. 

The same conclusion can be drawn by noting that a $B^0$ can resonate with a
positive pion to form an excited $B^+$, which we shall call $B^{**+}$ (to
distinguish it from the $B^*$, lying less than 50 MeV/$c^2$ above the $B$).
Similarly, a $\bar B^0$ can resonate with a negative pion to form a $B^{**-}$. 
The combinations $B^0 \pi^-$ and $\bar B^0 \pi^+$ are {\it exotic},~i.e., they
cannot be formed as quark-antiquark states.  No evidence for exotic resonances
exists. 

\subsection{Results from simulation}

We have asked the authors of a fragmentation Monte Carlo program to see if the
correlation between pions and neutral $B$ mesons is evident in their work.  The
result, based on $10^6$ events generated using ARIADNE and JETSET \cite{LUN},
shows a slight excess of the ``right-sign'' combinations $B^0 \pi^+$ over the
``wrong-sign'' combinations $B^0 \pi^-$. The ratio (right -- wrong)/(right +
wrong) varies from 0.17 for $M(B \pi) = 5.5$ GeV/$c^2$ to 0.27 for $M(B \pi) =
5.8 - 6.2$ GeV/$c^2$ (where there are fewer events).  No explicit resonances
were put into the simulation; their inclusion would strengthen the correlation.

\subsection{$B^{**}$ resonances}

The existence of a soft pion in $D^* \to D \pi$ decays \cite{NUS} has been a
key feature in tagging the presence of $D$ mesons since the earliest days of
charmed particles.  The mass of a $D^*$ is just large enough that the decays
$D^* \to D \pi$ can occur (except in the case of $D^0 \to \pi^- D^+$).  In
contrast, the $B^*$ is only 46 MeV above the $B$, so it cannot decay via
pion emission.  The lightest states which can decay to $B \pi$ and/or
$B^* \pi$ are P-wave resonances of a $b$ quark and an $\bar u$ or $\bar d$.
The expectations for masses of these states \cite{GRD,EHQ}, based on
extrapolation from the known $D^{**}$ resonances, are summarized in
Table~\ref{tab:res}.

\begin{table}
\begin{center}
\caption{\label{tab:res}P-wave resonances of a $b$ quark and a light ($\bar u$
or $\bar d$) antiquark} 
\medskip
\begin{tabular}{c c c} \hline
$J^P$  &    Mass     &  Allowed final \\
       & (GeV/$c^2$) &     state(s)   \\ \hline
$2^+$  & $\sim 5.77$ &  $B \pi,~B^* \pi$ \\
$1^+$  & $\sim 5.77$ &     $B^* \pi$  \\
$1^+$  & $ < 5.77$   &     $B^* \pi$  \\
$0^+$  & $ < 5.77$   &      $B \pi$   \\ \hline
\end{tabular}
\end{center}
\end{table}

The known $D^{**}$ resonances are a $2^+$ state around 2460 MeV/$c^2$, decaying
to $D \pi$ and $D^* \pi$, and a $1^+$ state around 2420 MeV/$c^2$, decaying to
$D^* \pi$.  These states are relatively narrow, probably because they decay
via a D-wave.  In addition, there are expected to be much broader (and probably
lower) $D^{**}$ resonances:  a $1^+$ state decaying to $D^* \pi$ and a $0^+$
state decaying to $D \pi$, both via S-waves.

\begin{figure}
% \vspace{3.5in}
\centerline{\epsfysize = 3.5in \epsffile {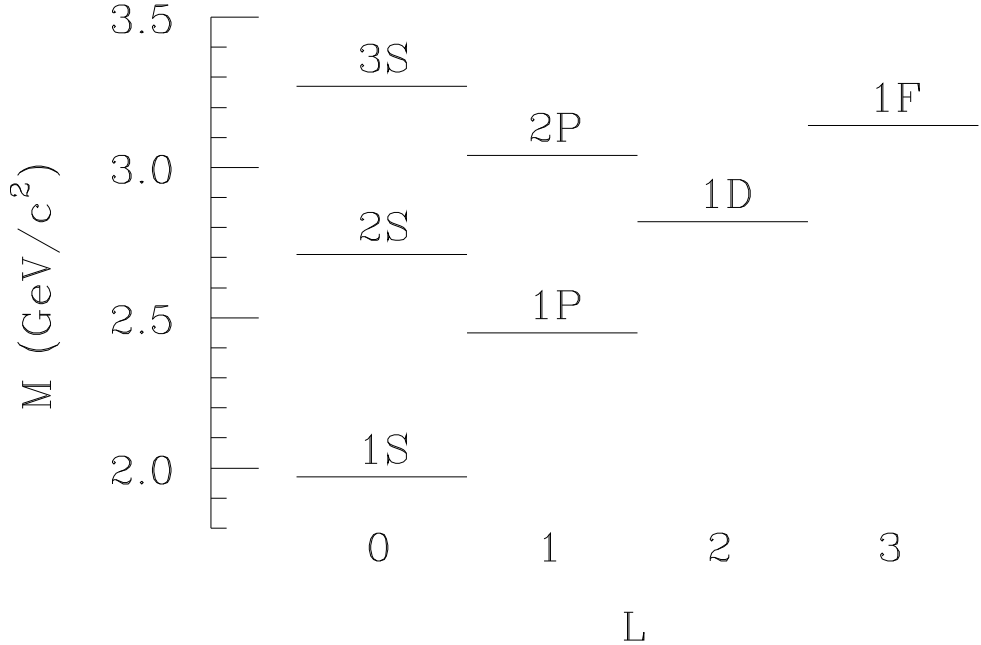}}
\caption{\label{fig:spe} Predicted spectrum \protect \cite{EHQ,QPC} of
nonstrange charmed meson resonances.  Observed states are labeled by a check
mark} 
\end{figure}

The expected spectrum of nonstrange charmed meson resonances is shown in
Fig.~\ref{fig:spe} \cite{EHQ,QPC}, as calculated in the potential of
Ref.~\cite{BT}. For strange states, one should add about 0.1 GeV, while for
$B$'s one should add about 3.32 GeV.  The predicted narrow $B^{**}$ resonances
lie at 5767 MeV $(2^+)$ and 5755 MeV ($1^+$).  These are the states in which
the light quark spin $s = 1/2$ and the orbital angular momentum $L = 1$ combine
to form a total light-quark angular momentum $j = 3/2$.  One also expects broad
$1^+$ and $0^+$ $B^{**}$ states with $j = 1/2$, decaying via S-waves. 

\subsection{The question of coherence}

As mentioned, one expects the $B^0 \bar B^0$ pair produced at the
$\Upsilon(4S)$ resonance to be in a state of charge-conjugation eigenvalue $C =
-1$.  The particle which decays at a time $t_1$ to a $CP$ eigenstate is then a
coherent superposition of $B^0$ and $\bar B^0$ when the particle produced in
association decays at a time $t_2$ to a tagging final state (e.g., to $D^{*+}
e^- \bar \nu_e$).  The fact that the time-dependent rate asymmetry is an
odd function of $t_1 - t_2$ is why one has to stretch out the decay region
using asymmetric kinematics.

On the other hand, in the high energy associated production of pairs of
$b$-flavored hadrons, one usually assumes no coherence between $B^0$ and $\bar
B^0$ on one side of the reaction when tagging on the other.  Thus, the decaying
state is assumed to be an incoherent {\it mixture} of $B^0$ and $\bar B^0$
with specific probabilities of each.

M. Gronau and I \cite{GRL} have proposed a way to test for coherence using
a density-matrix formalism.  One can measure the elements of the density
matrix using time-dependences of appearance of specific final states.

We work in a two-component basis labeled either by $B^0$ and $\bar B^0$ or,
more conveniently, by mass eigenstates.  In this last basis, in which
components of the density matrix are labeled by primed quantities, we
can denote an arbitrary coherent or incoherent state by the density
matrix
\begin{equation}
\rho = \frac{1}{2} \left( 1 + {\bf Q' \cdot \sigma} \right)~~~
\end{equation}
where $\sigma_i$ are the Pauli matrices.  The intensities for decays to states
of identified flavor can be written
\begin{equation}
I \left( \begin{array}{c} B^0 \\ \overline{B}^0 \\ \end{array} \right) =
\frac{1}{2} |A|^2 e^{-\Gamma t} \left[ 1 \pm Q_{\perp}' \cos (\Delta m t +
\delta) \right]~~~. 
\end{equation}
where
\begin{equation}
Q'_1 = Q'_\perp \cos \delta~~~,~~~~Q'_2 = Q'_\perp \sin \delta~~~.
\end{equation}
In order to measure $Q'_3$ one needs also to see decays to $CP$ eigenstates,
such as $J/\psi K_S$.  One then learns not only the components of the density
matrix, but also one of the angles of the unitarity triangle (such as
$\beta$).  Then, one can look at other final states to learn other angles.
For example, if penguin diagrams are not important in the decays $B \to
\pi \pi$, the final state $\pi^+ \pi^-$ provides information on the
angle $\alpha$.

\section{Conclusions}\label{sec:con}

The present knowledge about magnitudes and phases of CKM matrix elements
allows lots of ``wiggle room'' for inventive schemes.  Choosing among these
schemes will require progress on many fronts.  Among these, we have discussed
improved knowledge of meson decay constants, decays of neutral kaons, rare
$B$ decays, mixing of strange neutral $B$ mesons with their antiparticles,
and the observation of $CP$ violation in decays of neutral $B$ mesons.  The
identification of the initial flavor of a neutral $B$ meson may profit from
the study of hadrons produced in association with it, and we have described
ways in which our knowledge of such correlations may be improved.

\section{Acknowledgments}

I would like to thank several people for enjoyable collaborations on
aspects of the work presented here:  Geoff Harris on the work which permitted
several of the figures to be drawn, and Alex Nippe and Michael Gronau on the
topics mentioned in Section~\ref{sec:tag}.  Bill Badgett helped to prepare
Fig.~\ref{fig:mwmt}.  This work was supported in part by
the United States Department of Energy under Grant No. DE FG02 90ER40560. 

%%%%%%%%%%%%%%%%%%%%%%%%%%%
%
% Journal and other miscellaneous abbreviations for references
% World Scientific format for journals
\def \ap#1#2#3{{\it Ann. Phys. (N.Y.)} {\bf#1} (#3) #2}
\def \app#1#2#3{{\it Acta Physica Polonica} {\bf#1} (#3) #2}
\def \arnps#1#2#3{{\it Ann. Rev. Nucl. Part. Sci.} {\bf#1} (#3) #2}
\def \arns#1#2#3{{\it Ann. Rev. Nucl. Sci.} {\bf#1} (#3) #2}
\def \ba88{{\it Particles and Fields 3} (Proceedings of the 1988 Banff Summer
Institute on Particles and Fields), edited by A. N. Kamal and F. C. Khanna
(World Scientific, Singapore, 1989)}
\def \baphs#1#2#3{{\it Bull. Am. Phys. Soc.} {\bf#1} (#3) #2}
\def \be87{{\it Proceedings of the Workshop on High Sensitivity Beauty
Physics at Fermilab,} Fermilab, Nov. 11-14, 1987, edited by A. J. Slaughter,
N. Lockyer, and M. Schmidt (Fermilab, Batavia, IL, 1988)} 
\def \cn{Collaboration}
\def \cp89{{\it CP Violation,} edited by C. Jarlskog (World Scientific,
Singapore, 1989)} 
\def \dpf91{{\it The Vancouver Meeting - Particles and Fields '91}
(Division of Particles and Fields Meeting, American Physical Society,
Vancouver, Canada, Aug.~18-22, 1991), ed. by D. Axen, D. Bryman, and M. Comyn
(World Scientific, Singapore, 1992)}
\def \dpff{{\it The Fermilab Meeting - DPF 92} (Division of Particles and Fields
Meeting, American Physical Society, Fermilab, 10 -- 14 November, 1992), ed. by
C. H. Albright \ite~(World Scientific, Singapore, 1993)}
\def \hb87{{\it Proceeding of the 1987 International Symposium on Lepton and
Photon Interactions at High Energies,} Hamburg, 1987, ed. by W. Bartel
and R. R\"uckl (Nucl. Phys. B, Proc. Suppl., vol. 3) (North-Holland,
Amsterdam, 1988)}
\def \ib{{\it ibid.}~}
\def \ibj#1#2#3{{\it ibid.} {\bf#1} (#3) #2}
\def \ijmpa#1#2#3{{\it Int. J. Mod. Phys.} A {\bf#1} (#3) #2}
\def \ite{{\it et al.}}
\def \jpg#1#2#3{{\it J. Phys.} G {\bf#1} (#3) #2}
\def \ky85{{\it Proceedings of the International Symposium on Lepton and
Photon Interactions at High Energy,} Kyoto, Aug.~19-24, 1985, edited by M.
Konuma and K. Takahashi (Kyoto Univ., Kyoto, 1985)} 
\def \lat90{{\it Results and Perspectives in Particle Physics} (Proceedings of
Les Rencontres de Physique de la Vallee d'Aoste [4th], La Thuile, Italy, Mar.
18-24, 1990), edited by M. Greco (Editions Fronti\`eres, Gif-Sur-Yvette, France,
1991)}
\def \lg91{International Symposium on Lepton and Photon Interactions, Geneva,
Switzerland, July, 1991}
\def \lkl87{{\it Selected Topics in Electroweak Interactions} (Proceedings of 
the Second Lake Louise Institute on New Frontiers in Particle Physics, 15 --
21 February, 1987), edited by J. M. Cameron \ite~(World Scientific, Singapore,
1987)}
\def \mpla #1#2#3{{\it Mod. Phys. Lett.} A {\bf#1} (#3) #2}
\def \nc#1#2#3{{\it Nuovo Cim.} {\bf#1} (#3) #2}
\def \np#1#2#3{{\it Nucl. Phys.} {\bf#1} (#3) #2}
\def \pl#1#2#3{{\it Phys. Lett.} {\bf#1} (#3) #2}
\def \plb#1#2#3{{\it Phys. Lett.} B {\bf#1} (#3) #2}
\def \ppnp#1#2#3{{\it Prog. Part. Nucl. Phys.} {\bf#1} (#3) #2}
\def \pr#1#2#3{{\it Phys. Rev.} {\bf#1} (#3) #2}
\def \prd#1#2#3{{\it Phys. Rev.} D {\bf#1} (#3) #2}
\def \prl#1#2#3{{\it Phys. Rev. Lett.} {\bf#1} (#3) #2}
\def \prp#1#2#3{{\it Phys. Rep.} {\bf#1} (#3) #2}
\def \ptp#1#2#3{{\it Prog. Theor. Phys.} {\bf#1} (#3) #2}
\def \rmp#1#2#3{{\it Rev. Mod. Phys.} {\bf#1} (#3) #2}
\def \rp#1{~~~~~\ldots\ldots{\rm rp~}{#1}~~~~~}
\def \si90{25th International Conference on High Energy Physics, Singapore,
Aug. 2-8, 1990, Proceedings edited by K. K. Phua and Y. Yamaguchi (World
Scientific, Teaneck, N. J., 1991)}
\def \slac75{{\it Proceedings of the 1975 International Symposium on
Lepton and Photon Interactions at High Energies,} Stanford University, Aug.
21-27, 1975, edited by W. T. Kirk (SLAC, Stanford, CA, 1975)} 
\def \slc87{{\it Proceedings of the Salt Lake City Meeting} (Division of
Particles and Fields, American Physical Society, Salt Lake City, Utah, 1987),
ed. by C. DeTar and J. S. Ball (World Scientific, Singapore, 1987)}
\def \smass82{{\it Proceedings of the 1982 DPF Summer Study on Elementary
Particle Physics and Future Facilities}, Snowmass, Colorado, edited by R.
Donaldson, R. Gustafson, and F. Paige (World Scientific, Singapore, 1982)}
\def \smass90{{\it Research Directions for the Decade} (Proceedings of the
1990 DPF Snowmass Workshop), edited by E. L. Berger (World Scientific,
Singapore, 1991)}
\def \stone{{\it B Decays}, edited by S. Stone (Singapore:  World
Scientific, 1994)}
\def \tasi90{{\it Testing the Standard Model} (Proceedings of the 1990
Theoretical Advanced Study Institute in Elementary Particle Physics),
edited by M. Cveti\v{c} and P. Langacker (World Scientific, Singapore, 1991)}
\def \yaf#1#2#3#4{{\it Yad. Fiz.} {\bf#1} (#3) #2 [Sov. J. Nucl. Phys. {\bf #1}
 (#3) #4]}
\def \zhetf#1#2#3#4#5#6{{\it Zh. Eksp. Teor. Fiz.} {\bf #1} (#3) #2 [Sov.
Phys. - JETP {\bf #4} (#6) #5]}
\def \zhetfl#1#2#3#4{{\it Pis'ma Zh. Eksp. Teor. Fiz.} {\bf #1} (#3) #2 [JETP
Letters {\bf #1} (#3) #4]}
\def \zp#1#2#3{{\it Zeit. Phys.} {\bf#1} (#3) #2}
\def \zpc#1#2#3{{\it Zeit. Phys.} C {\bf#1} (#3) #2}
%%%%%%%%%%%%%%%%%%%%%%%%%%%%%%%%%%%%%%%%%%%%%%%%%%%

\end{document}